\documentclass{emulateapj}

\usepackage{lineno}

\shorttitle{Spatially resolving MGRO J2019+37 with VERITAS}
\shortauthors{Aliu et al.}

\begin{document}

\title{Spatially Resolving the Very High Energy emission from MGRO J2019+37 with VERITAS}

\author{
E.~Aliu\altaffilmark{1,*},
T.~Aune\altaffilmark{2},
B.~Behera\altaffilmark{3},
M.~Beilicke\altaffilmark{4},
W.~Benbow\altaffilmark{5},
K.~Berger\altaffilmark{6},
R.~Bird\altaffilmark{7},
A.~Bouvier\altaffilmark{8},
J.~H.~Buckley\altaffilmark{4},
V.~Bugaev\altaffilmark{4},
M.~Cerruti\altaffilmark{5},
X.~Chen\altaffilmark{9,3},
L.~Ciupik\altaffilmark{10},
M.~P.~Connolly\altaffilmark{11},
W.~Cui\altaffilmark{12},
J.~Dumm\altaffilmark{13},
V.~V.~Dwarkadas\altaffilmark{14},
M.~Errando\altaffilmark{1},
A.~Falcone\altaffilmark{15},
S.~Federici\altaffilmark{3,9},
Q.~Feng\altaffilmark{12},
J.~P.~Finley\altaffilmark{12},
H.~Fleischhack\altaffilmark{3},
P.~Fortin\altaffilmark{5},
L.~Fortson\altaffilmark{13},
A.~Furniss\altaffilmark{8},
N.~Galante\altaffilmark{5},
G.~H.~Gillanders\altaffilmark{11},
E.~V.~Gotthelf\altaffilmark{16},
S.~Griffin\altaffilmark{17},
S.~T.~Griffiths\altaffilmark{18},
J.~Grube\altaffilmark{10},
G.~Gyuk\altaffilmark{10},
D.~Hanna\altaffilmark{17},
J.~Holder\altaffilmark{6},
G.~Hughes\altaffilmark{3},
T.~B.~Humensky\altaffilmark{19},
C.~A.~Johnson\altaffilmark{8},
P.~Kaaret\altaffilmark{18},
O.~Kargaltsev\altaffilmark{20},
M.~Kertzman\altaffilmark{21},
Y.~Khassen\altaffilmark{7},
D.~Kieda\altaffilmark{22},
F.~Krennrich\altaffilmark{23},
M.~J.~Lang\altaffilmark{11},
A.~S~Madhavan\altaffilmark{23},
G.~Maier\altaffilmark{3},
S.~McArthur\altaffilmark{24},
A.~McCann\altaffilmark{25},
J.~Millis\altaffilmark{26,27},
P.~Moriarty\altaffilmark{27},
R.~Mukherjee\altaffilmark{1},
D.~Nieto\altaffilmark{19},
A.~O'Faol\'{a}in de Bhr\'{o}ithe\altaffilmark{7},
R.~A.~Ong\altaffilmark{2},
A.~N.~Otte\altaffilmark{28},
D.~Pandel\altaffilmark{29},
N.~Park\altaffilmark{24,*},
M.~Pohl\altaffilmark{9,3},
A.~Popkow\altaffilmark{2},
H.~Prokoph\altaffilmark{3},
J.~Quinn\altaffilmark{7},
K.~Ragan\altaffilmark{17},
J.~Rajotte\altaffilmark{17},
L.~C.~Reyes\altaffilmark{30},
P.~T.~Reynolds\altaffilmark{31},
G.~T.~Richards\altaffilmark{28},
E.~Roache\altaffilmark{5},
M.~Roberts\altaffilmark{32},
G.~H.~Sembroski\altaffilmark{12},
K.~Shahinyan\altaffilmark{13},
A.~W.~Smith\altaffilmark{22},
D.~Staszak\altaffilmark{17},
I.~Telezhinsky\altaffilmark{9,3},
J.~V.~Tucci\altaffilmark{12},
J.~Tyler\altaffilmark{17},
S.~Vincent\altaffilmark{3},
S.~P.~Wakely\altaffilmark{24},
A.~Weinstein\altaffilmark{23},
R.~Welsing\altaffilmark{3},
A.~Wilhelm\altaffilmark{9},
D.~A.~Williams\altaffilmark{8},
B.~Zitzer\altaffilmark{34}
}

\altaffiltext{1}{Department of Physics and Astronomy, Barnard College, Columbia University, NY 10027, USA}
\altaffiltext{2}{Department of Physics and Astronomy, University of California, Los Angeles, CA 90095, USA}
\altaffiltext{3}{DESY, Platanenallee 6, 15738 Zeuthen, Germany}
\altaffiltext{4}{Department of Physics, Washington University, St. Louis, MO 63130, USA}
\altaffiltext{5}{Fred Lawrence Whipple Observatory, Harvard-Smithsonian Center for Astrophysics, Amado, AZ 85645, USA}
\altaffiltext{6}{Department of Physics and Astronomy and the Bartol Research Institute, University of Delaware, Newark, DE 19716, USA}
\altaffiltext{7}{School of Physics, University College Dublin, Belfield, Dublin 4, Ireland}
\altaffiltext{8}{Santa Cruz Institute for Particle Physics and Department of Physics, University of California, Santa Cruz, CA 95064, USA}
\altaffiltext{9}{Institute of Physics and Astronomy, University of Potsdam, 14476 Potsdam-Golm, Germany}
\altaffiltext{10}{Astronomy Department, Adler Planetarium and Astronomy Museum, Chicago, IL 60605, USA}
\altaffiltext{11}{School of Physics, National University of Ireland Galway, University Road, Galway, Ireland}
\altaffiltext{12}{Department of Physics, Purdue University, West Lafayette, IN 47907, USA }
\altaffiltext{13}{School of Physics and Astronomy, University of Minnesota, Minneapolis, MN 55455, USA}
\altaffiltext{14}{Department of Astronomy and Astrophysics, University of Chicago, Chicago, IL, 60637}
\altaffiltext{15}{Department of Astronomy and Astrophysics, 525 Davey Lab, Pennsylvania State University, University Park, PA 16802, USA}
\altaffiltext{16}{Columbia Astrophysics Laboratory, Columbia University, New York, NY 10027}
\altaffiltext{17}{Physics Department, McGill University, Montreal, QC H3A 2T8, Canada}
\altaffiltext{18}{Department of Physics and Astronomy, University of Iowa, Van Allen Hall, Iowa City, IA 52242, USA}
\altaffiltext{19}{Physics Department, Columbia University, New York, NY 10027, USA}
\altaffiltext{20}{Department of Physics, The George Washington University, Washington, DC 20052, USA}
\altaffiltext{21}{Department of Physics and Astronomy, DePauw University, Greencastle, IN 46135-0037, USA}
\altaffiltext{22}{Department of Physics and Astronomy, University of Utah, Salt Lake City, UT 84112, USA}
\altaffiltext{23}{Department of Physics and Astronomy, Iowa State University, Ames, IA 50011, USA}
\altaffiltext{24}{Enrico Fermi Institute, University of Chicago, Chicago, IL 60637, USA}
\altaffiltext{25}{Kavli Institute for Cosmological Physics, University of Chicago, Chicago, IL 60637, USA}
\altaffiltext{26}{Department of Physics, Anderson University, 1100 East 5th Street, Anderson, IN 46012}
\altaffiltext{27}{Department of Life and Physical Sciences, Galway-Mayo Institute of Technology, Dublin Road, Galway, Ireland}
\altaffiltext{28}{School of Physics and Center for Relativistic Astrophysics, Georgia Institute of Technology, 837 State Street NW, Atlanta, GA 30332-0430}
\altaffiltext{29}{Department of Physics, Grand Valley State University, Allendale, MI 49401, USA}
\altaffiltext{30}{Physics Department, California Polytechnic State University, San Luis Obispo, CA 94307, USA}
\altaffiltext{31}{Department of Applied Physics and Instrumentation, Cork Institute of Technology, Bishopstown, Cork, Ireland}
\altaffiltext{32}{Eureka Scientific Inc., 2452 Delmer Street, Suite 100, Oakland, CA 94602-3017, USA}
\altaffiltext{33}{New York Unversity of Abu Dhabi, United Arab Emirates}
\altaffiltext{34}{Argonne National Laboratory, 9700 S. Cass Avenue, Argonne, IL 60439, USA}


\altaffiltext{*}{Address correspondence to E.~Aliu or N.~Park \\
E-mail: ealiu@astro.columbia.edu or nahee@uchicago.edu}


\begin{abstract}
We present very high energy (VHE) imaging of MGRO~J2019$+37$ obtained with the VERITAS observatory. The bright extended ($\sim2^{\circ}$) unidentified Milagro source is located towards the rich star formation region Cygnus-X. MGRO~J2019$+37$ is resolved into two VERITAS sources. The faint point-like source VER~J2016$+$371 overlaps CTB~87, a filled-center remnant (SNR) with no evidence of a supernova remnant shell at the present time. Its spectrum is well fit in the $0.65 - 10$~TeV energy range by a power-law model with photon index $2.3\pm0.4$.
VER~J2019$+$368 is a bright extended ($\sim1^{\circ}$) source, that likely accounts for the bulk of the Milagro emission and is notably coincident with PSR~J2021$+$3651 and the star formation region Sh~2$-$104. Its spectrum in the range $1-30$~TeV is well fit with a power-law model of photon index $1.75\pm0.3$, among the hardest values measured in the VHE band, comparable to that observed near Vela-X. We explore the unusual spectrum and morphology in the radio and X-ray bands to constrain possible emission mechanisms for this source.

\end{abstract}


\section{Introduction}

High-mass star formation (and death) has long been associated with the acceleration of very high energy (VHE; $>$100 GeV) particles~\citep{ginzburg64} and gamma-ray emission \citep{montmerle79,kaaret96}. There is clear evidence of VHE particle acceleration in the  products of stellar death such as pulsars, supernova remnant (SNR) shells and pulsar wind nebulae (PWNe). Star-forming regions produce copious kinetic power in other forms, such as winds from Wolf-Rayet and OB stars. Regions with stellar winds, from single, binary, or collections of stars, have been suggested as possible VHE gamma-ray emitting sites, but as of today,~observational evidence of this is still very scarce~\citep{lemoinegoumard11}. 

The  Cygnus-X region is one of the richest known regions of star formation in the Galaxy.  It is also close by (at only ~1.4 kpc) and is, therefore, an excellent laboratory to study high-energy particle acceleration related to high-mass star formation and death. Because there are several spiral arms in the same direction, care must be exercised in relating any individual source to Cygnus-X. The Milagro sky survey identified several bright and extended VHE gamma-ray sources in the general direction of Cygnus~\citep{abdo07b}. However, each  Milagro source has multiple possible counterparts at lower energies which complicates unambiguous associations. 

\begin{figure*}[t]
\epsscale{0.8}
\plotone{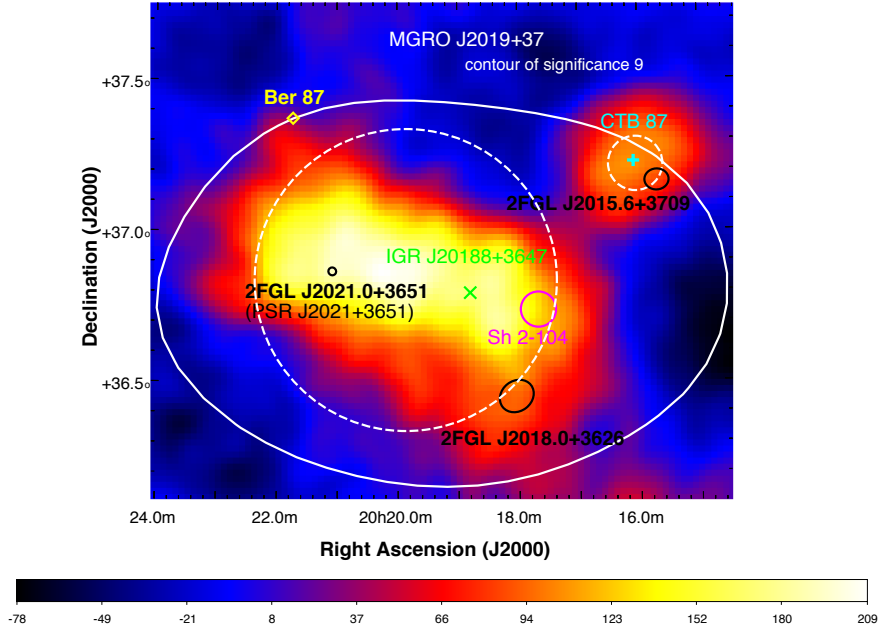}
\caption{VHE gamma-ray excess map of the MGRO J2019+37 region as observed by VERITAS above 600 GeV. The color bar indicates the number of excess events within a search radius of 0$^{\circ}$.23, which corresponds to the extended source search analysis described in the text. The shift between red and blue color scale occurs at the 3$\sigma$ level. Regions used for spectral analysis of VER J2016+371 and J2019+368 are defined by white dashed circles. The locations of possible counterparts are marked using different colors. The contour of significance 9$\sigma$ of MGRO J2019+37 is overlaid in white. \label{fig1}}
\end{figure*}

\begin{figure*}[t]
\epsscale{0.8}
\plotone{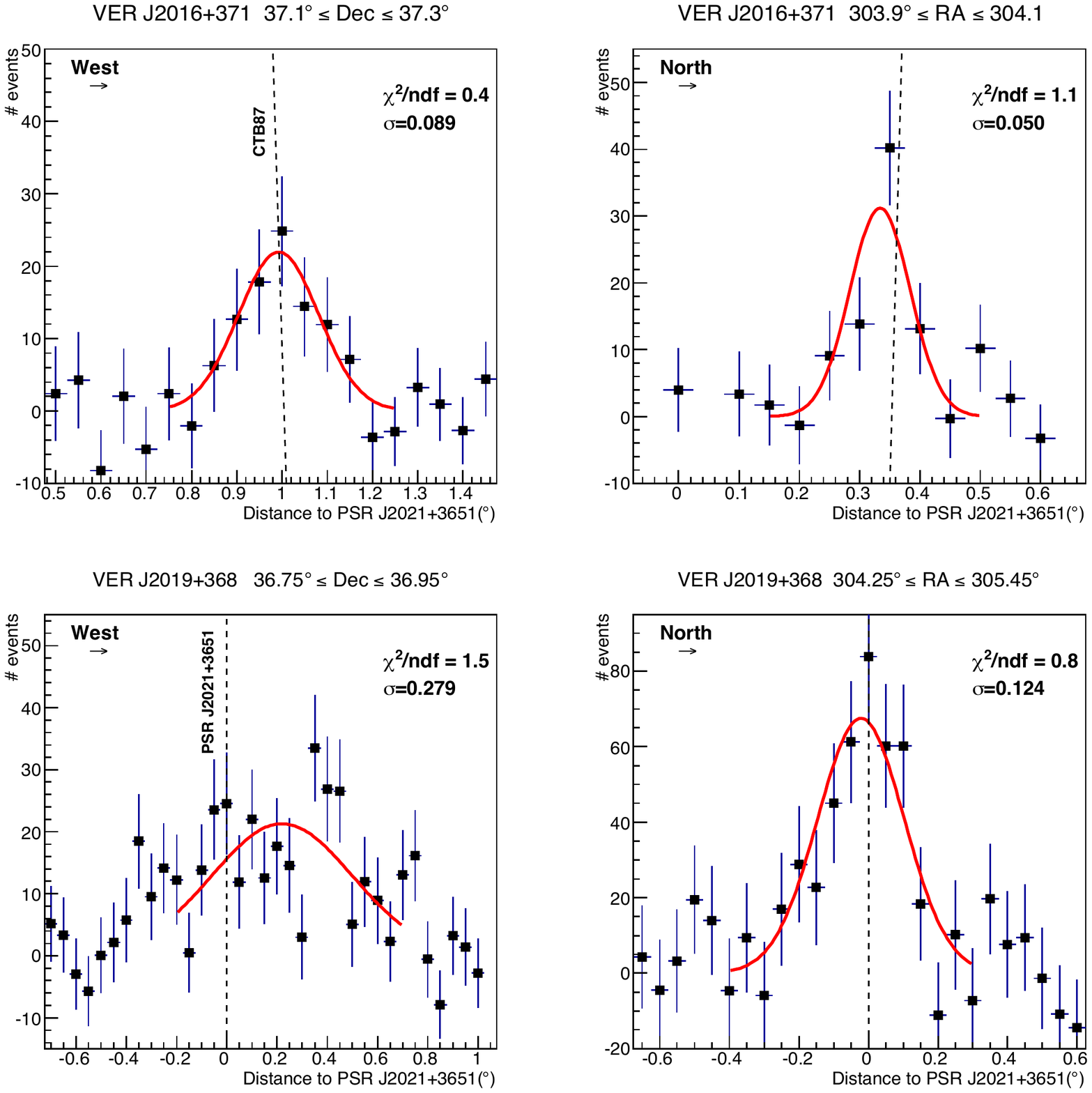}
\caption{ Slices in the uncorrelated excess maps of  the new VHE sources. The width of the slice is indicated on each panel. The slices are centered on the best fit position of each VHE source. The direction of the slices follows right ascension. The top panels show the slices for VER J2016+371 and the peak position of the 1420 MHz emission of CTB 87~\citep{kothes03} is indicated with a dashed line. The bottom panels correspond to VER J2019+368 and the dashed line shows the position of the pulsar PSR J2021+3651.    \label{fig2}}
\end{figure*}

MGRO J2019+37, with a measured flux of about 80\% of the Crab Nebula flux at 20 TeV~\citep{abdo07b}, is the brightest Milagro source in the region. Since its discovery, the nature of MGRO J2019+37 has been the subject of studies and speculation, yet it remains unknown. Its bright inner region has an extent of about 1$^{\circ}$ and overlaps with several SNRs, H{\sc ii} regions, Wolf-Rayet stars, high-energy gamma-ray (HE; $>$100 MeV) sources, and a hard X-ray transient. A tentative association with the young energetic radio and gamma-ray pulsar PSR J2021+3651 and its nebula, SNR G75.1+0.2, has been suggested~\citep{abdo07b}. There is extended X-ray and radio emission associated with the pulsar, but the size (less than $10'$, \citet{roberts08}) is significantly smaller than the VHE source measured with Milagro. \cite{paredes09} suggested that this pulsar alone is not able to power the whole emission of MGRO J2019+37 because the time required for the electrons to diffuse and to fill a region of 1$^{\circ}$ (at an uncertain distance of 2 to 10 kpc) is larger than their cooling time. The same authors performed deep radio and near-infrared surveys to find other potential counterparts and proposed the massive star-forming region associated with the H{\sc ii} region Sharpless 104 (Sh 2-104) as a possible contributor to the VHE emission through wind collisions or interactions of protostar jets with the surrounding medium (e.g.\ \cite{torres04}). Particle acceleration in shocks driven by the winds from the Wolf-Rayet stars in the young cluster Ber 87 in the Cyg OB1 association has also been proposed as an origin of the VHE gamma rays \citep{bednarek07}. 

Several VHE instruments have reported results on the region near MGRO J2019+37 at energies below 10~TeV.  Relatively short observations with the imaging atmospheric Cherenkov telescopes (IACTs) MAGIC and VERITAS led to upper limits consistent with the Milagro source being extended and hard \citep{bartko08, kieda08}. \citet{bartoli12} recently reported a non-detection based on data from the air-shower array ARGO-YBJ and concluded that the source could be variable. Only the Tibet Air Shower array has confirmed the detection of an extended VHE source from the same direction, with a statistical significance of 5.8 standard deviations (5.8$\sigma$)~\citep{amenomori08}. 

In this paper we report on new and deeper observations of the region around MGRO J2019+37 made with VERITAS.  The new observations provide much better angular resolution than Milagro and better sensitivity than any of the previous VHE measurements. These observations enable us to map the VHE emission in an attempt to better understand its physical origin. The instrument and observations are described in \S~2, while analysis and results can be found in \S~3.  A multi-wavelength analysis of the possible counterparts to this emission and a general discussion are presented in \S~4. A short summary and conclusions are drawn in \S~5.

 \section{VHE observations}

VERITAS is an array of four IACTs designed for observations of astrophysical objects in the energy range from 100 GeV to above 10 TeV.~The instrument angular resolution (68\% containment) reaches $0^{\circ}.08$ per event and its sensitivity for a point-source is 1\% of the steady Crab Nebula flux above 300 GeV for a 5$\sigma$ detection within 25 hours of observation at a 20$^{\circ}$ zenith angle. A review of the detector is given by~\cite{holder06, holder08}.

The first pointed observations of the region around MGRO J2019+37 with VERITAS took place in November 2006, during its commissioning phase, when the array had only two telescopes.~The accumulated 10 hours of exposure did not lead to any detection \citep{kieda08}. The region was re-observed during the VERITAS survey of the Cygnus region from 2007 to 2008 using the full array~\citep{weinstein09}.~Analysis of the survey data with an effective on-source exposure of 7 hours revealed a hint of a VHE gamma-ray signal within the large extension of the Milagro source. 

A dedicated observation of the region took place between April 2010 and December 2010, resulting in 70 hours available for analysis after data quality selection. These observations were taken using the so-called wobble mode method, in which the source is offset by a small angular distance from the center of the field of view, alternating between 20 minute runs in the four cardinal directions on the sky. This method ensures a simultaneous background estimate in each run. The size of the offset was decided based on the expected very large extent of the source and to have the maximum coverage to all objects in the field. 
The first 46 hours of the observations used an offset of $0^{\circ}.7$ around PSR J2021+3651, while 21 hours of the observations used an offset of $0^{\circ}.6$ around the hard X-ray transient IGR J20188+3657, located near the centroid of MGRO J2019+37. An additional 4 hours were also taken with offsets of $0^{\circ}.5$ from the SNR CTB 87 to increase the exposure on the source. In this study, we only present data from these dedicated observations. The zenith angle of the selected data set ranges from $5^{\circ}$ to $35^{\circ}$.

\section{VHE gamma-ray analysis and results}
\subsection{Analysis}
The results presented here were generated using one of the standard VERITAS event reconstruction packages, similar to the scheme described in~\cite{acciari08}.~All results were then verified by using a second independent software package described in~\cite{cogan08}.~Air shower images fully contained in the individual telescope cameras are parameterized using the standard Hillas moment analysis~\citep{hillas85}. If the air shower image is not fully contained, a simple log-likelihood fitting of the image to a two-dimensional ellipse, which is generally accepted as a good representation of a gamma-ray shower~\citep{hillas85}, was applied in order to better estimate the shower parameters. The combination of both methods leads to more accurate reconstructions of off-axis events and a more than 30\% increase of the effective area above 5 TeV, and thus to an improved sensitivity. 

The spectral index of the gamma-ray source in the region, likely the counterpart of MGRO J2019+37, is expected to be hard ($\Gamma\sim2$) based on the previous result of~\cite{kieda08}. However, the number of potential sources and their extensions are unknown. Therefore, we define two sets of selection criteria, for point sources and for extended sources, both optimized for weak emission with a hard spectral index. For the point source search, we use an integration radius of  $\theta_{int}=0^{\circ}.089$, and for the extended source search, we use $\theta_{int}=0^{\circ}.23$.
The optimized cuts require a hard cut on the image intensity, of 225 photoelectrons, yielding a mean energy threshold of $\sim$600 GeV. 

For the background estimation of the sky image, the ring background method~\citep{aharonian05} was chosen with a background radius of 0$^{\circ}$.7, to avoid possible gamma-ray contamination due to the spatial extent of the source.
The reflected region background method~\citep{aharonian01} was used for the spectral analysis. Regions around stars with a magnitude less than 5.0 and the potential VHE sources PSR J2021+3651 and CTB 87 were excluded from the background estimation.

Trials have been conservatively estimated by tiling the area containing the Milagro source with $0^{\circ}.04$ square search bins, which yields 2500 trials (1250 for each set of cuts) in the center region.


\subsection{Results}
The VERITAS excess map of the MGRO J2019+37 region is shown in Fig.~\ref{fig1}, for the extended source search described previously.
~The map reveals at least two separate VHE emission regions where a single source had been reported by Milagro. Table~\ref{table:analysis result} summarizes the details of the analysis result for each emission region.

\begin{figure}[t] 
\epsscale{1.3}
\plotone{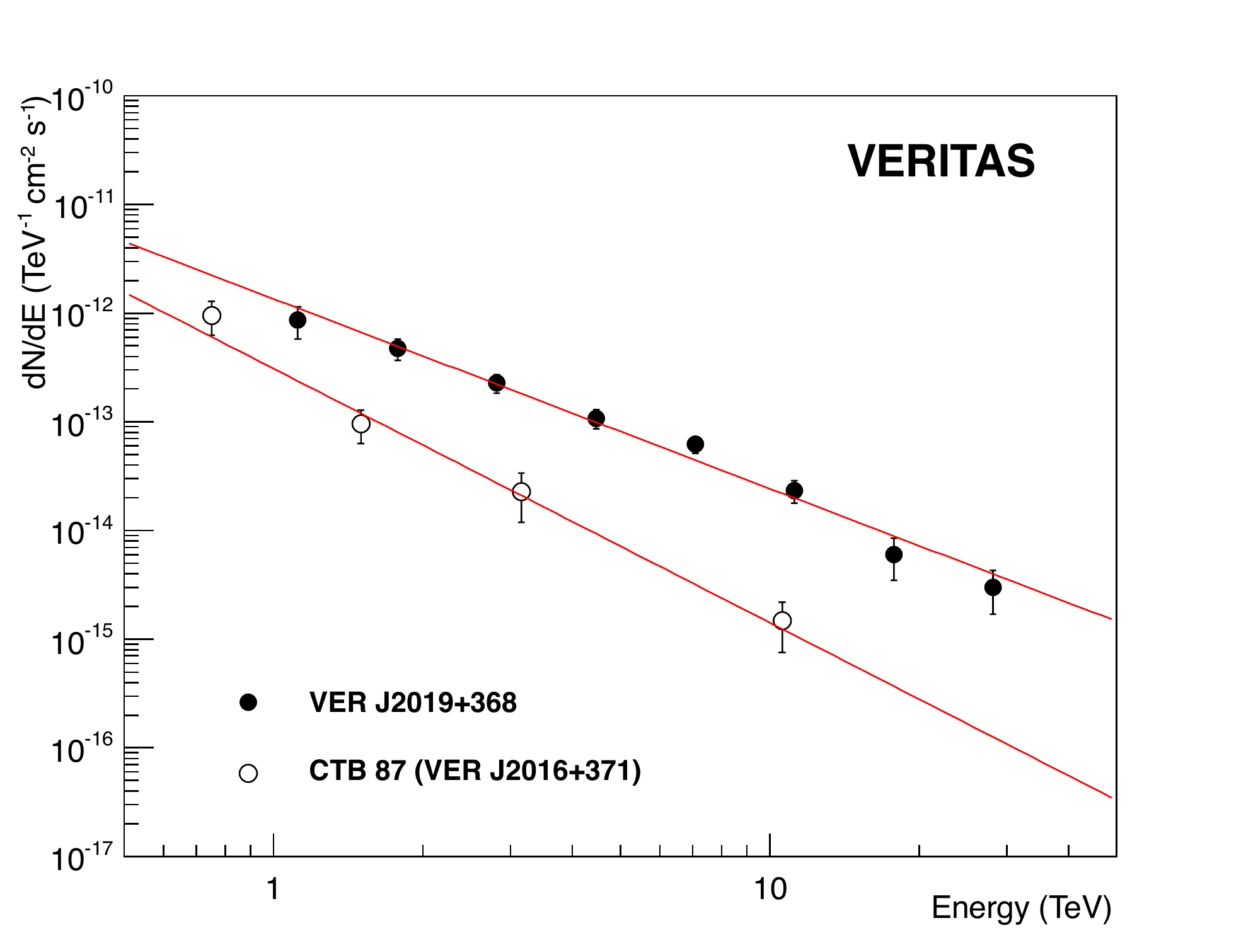}
\caption{Differential energy spectrum of VER J2016+371/CTB 87 and VER J2019+368 as measured by VERITAS. The event excess in each bin have a statistical significance of at least 2$\sigma$.  \label{fig3}}
\end{figure}

One of the regions is located in the northwest of the field-of-view (FOV), where the gamma-ray excess is detected at a post-trials significance of 5.8$\sigma$ using the smaller search radius. This source is point-like to VERITAS and its best fit position is found to be $\alpha_{J2000}= 20^h16^m2^s\pm3^s_{stat}$, $\delta_{J2000}=37^{\circ}11{\arcmin}52{\arcsec}\pm40{\arcsec}_{stat}$ and, hence, it is named VER J2016+371. This position is obtained by fitting a two-dimensional symmetric Gaussian ($\sigma_x = \sigma_y$) function to the uncorrelated excess map. The systematic uncertainty in the measurement is well below $\sim50\arcsec$.  The position of this new VHE source is consistent with the peak of the radio SNR CTB 87, less than $1{\arcmin}$ away~\citep{kothes03}. The top panel in Fig.~\ref{fig2} illustrates these results as well by showing the slices of the uncorrelated excess events of VER J2016+371. The directions of the slices are chosen along right ascension (R.A.) and perpendicular to R.A., while the widths of the slices are chosen to be  0$^{\circ}$.2.   
The energy spectrum of the VHE emission is obtained from a circular region of radius 0$^{\circ}$.09 centered on the nominal position of the SNR. 
This spectrum, shown in Fig.~\ref{fig3}, is well described by a power-law (PL) model ($\chi^{2}/$dof=1.82/2) that extends from 650 GeV up to 10 TeV with a photon index of $\Gamma=2.3 \pm 0.3_{stat} \pm 0.2_{sys}$ and a differential flux at 1 TeV of $(3.1 \pm 0.9_{stat} \pm 0.6_{sys})\times10^{-13}$ TeV$^{-1}$ cm$^{-2}$s$^{-1}$.~Using these best fit parameters, an integrated energy flux of  ($8.2 \pm 3.4_{stat} \pm 2.9_{sys})\times10^{-13}$ erg cm$^{-2}$s$^{-1}$ between 1 and 10 TeV is obtained. 

\begin{table*}[t]
\caption{ Analysis Results }
\centering
\begin{tabular}{ c c c c c c }
\hline\hline
Source name & $\it{On}$\footnote{$\it{On}$ is number of events in the source region} & $\it{Off}$\footnote{$\it{Off}$ is number of events in the background region} & $\alpha$\footnote{ $\alpha$ is defined by the ratio of the exposure of the source region to the exposure of background region.}  & Excess & Significance($\sigma$)\footnote{Significance shown here is before accounting for the trial factor. }  \\ [0.5ex]
\hline
VER J2016+371 & 126 & 317 & 0.181 & 69 & 7.0 \\
VER J2019+368 & 814 & 3656 & 0.160 & 228 & 8.2 \\ [1ex]
\hline
\end{tabular}
\label{table:analysis result}
\end{table*}

~The other emission region lies in the center of the map in Fig.~\ref{fig1}. The result of the search with the larger radius shows emission in a region about one degree in extent and elongated along the R.A. axis, which is detected at a post-trials significance of 7.2$\sigma$. The centroid and extension of the emission were estimated by fitting a two-dimensional asymmetric Gaussian ($\sigma_x\neq\sigma_y$) function convolved with the VERITAS PSF to the uncorrelated excess event map. The centroid is located at $\alpha_{J2000}=20^h19^m25^s\pm72^s_{stat}$, $\delta_{J2000}=36^{\circ}48{\arcmin}14{\arcsec}\pm58{\arcsec}_{stat}$ and, thus, labeled as VER J2019+368. From the fitting, the 1$\sigma$ angular extension of the emission was estimated to be $0^{\circ}.34\pm0^{\circ}.03_{stat}$ along the major axis, and $0^{\circ}.13\pm0^{\circ}.02_{stat}$ along the minor axis with the orientation angle of 71$^{\circ}$ east of north. 
~The extended emission region contains PSR J2021+3651 and its PWN, the H{\sc ii} region Sh 2-104 and the hard X-ray transient IGR J20188+3657, which are potential counterparts of the observed emission.  
To evaluate the possibility  of various source contributions to VER J2019+368 we did a morphology test. We extracted the profile of  the uncorrelated gamma-ray excess counts along the R.A. axis, see bottom panel of Fig~\ref{fig2}, restricted to 0$^{\circ}$.2 in declination, and fitted several possible VHE morphologies, including a single extended source and a superposition of point sources. All fits resulted in similar reduced $\chi^2$ values, mainly due to limited statistics. Either a larger data set or more sensitive reconstruction techniques, or both, are necessary to determine the morphology better.

\begin{figure}[t]
\epsscale{1.35}
\plotone{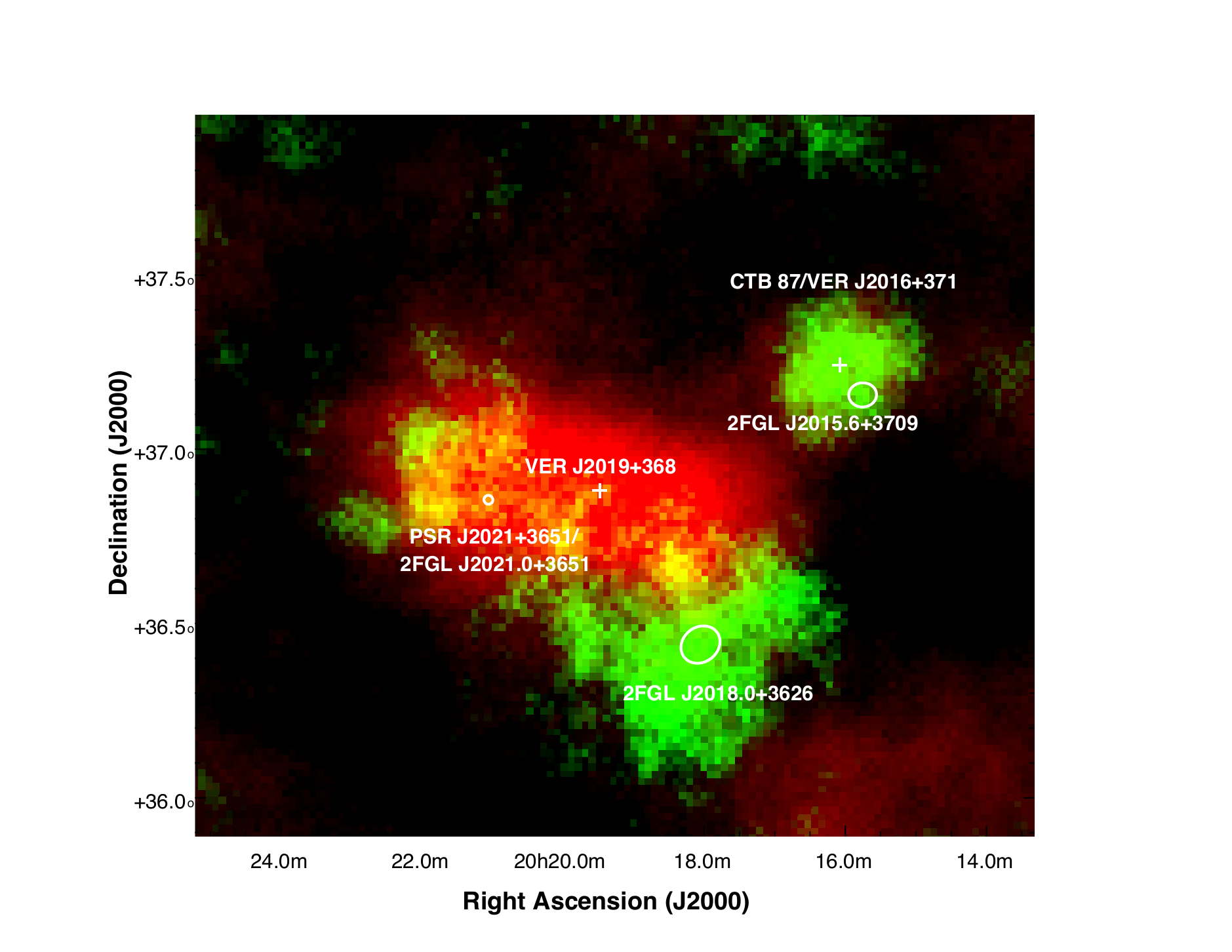}
\caption{VHE gamma-ray excess maps of the MGRO J2019+37 region as observed with VERITAS in two different energy bands. The high energy band is above 1 TeV (red) while the low energy band is between 600 GeV and 1 TeV (green). The number of excess events in the maps has been obtained using a search radius of 0$^{\circ}$.23, which corresponds to the extended source search analysis described in the text. The change between the red and black in the color scale takes place at the 4$\sigma$ level, while between green and black is at the 2$\sigma$ level. \label{fig4}}
\end{figure}

The energy spectrum for VER J2019+368 is estimated from a circular region of 0$^{\circ}$.5 radius centered on the best fit position. The resulting spectrum, shown in Fig.~\ref{fig3}, extends from 1 to 30 TeV and is well fit by a PL model ($\chi^{2}/dof=5.79/6$) with a hard photon index of $\Gamma=1.75\pm0.08_{stat}\pm0.2_{sys}$ and a differential flux at 5 TeV of $(8.1\pm0.7_{stat}\pm1.6_{sys})\times10^{-14}$ TeV$^{-1}$ cm$^{-2}$s$^{-1}$.~Assuming these parameters from the fit, the 1-10 TeV integrated energy flux is estimated to be $(6.7 \pm 0.5_{stat} \pm 1.2_{sys})\times10^{-12}$ erg cm$^{-2}$s$^{-1}$. We also attempted to fit alternative spectral models (such as a curved PL and cut-off PL model) but they did not provide better fits. The study of the energy dependent morphology of the emission in two separate energy bands, below 1 TeV, and above 1 TeV,  supports the lack of any statistically significant spectral points below 1 TeV.~The excess maps for each energy band show evidence for different centroid positions, see Fig.~\ref{fig4}.~Above 1 TeV, a strong emission (at the level of 9$\sigma$) with a best fit location statistically compatible with that of VER J2019+368 is observed. Below 1 TeV, there are indications (at the level of 3$\sigma$) of emission offset by about 0.5 degrees in the direction of the unidentified gamma-ray source 2FGL J2018.0+3626.

\section{Multiwavelength properties, interpretation and discussion}
Both VHE emitting regions coincide with non-thermal emission detected in radio, X-rays and HE gamma rays. In the following sections we examine in detail the locations, morphologies and spectral properties of these low energy counterparts in order to be able to establish the connection with the VHE emission and its origin.  

\subsection{ VER J2016+371, the SNR CTB 87 and their surroundings }

\begin{figure} [t]
\epsscale{1.2}
\plotone{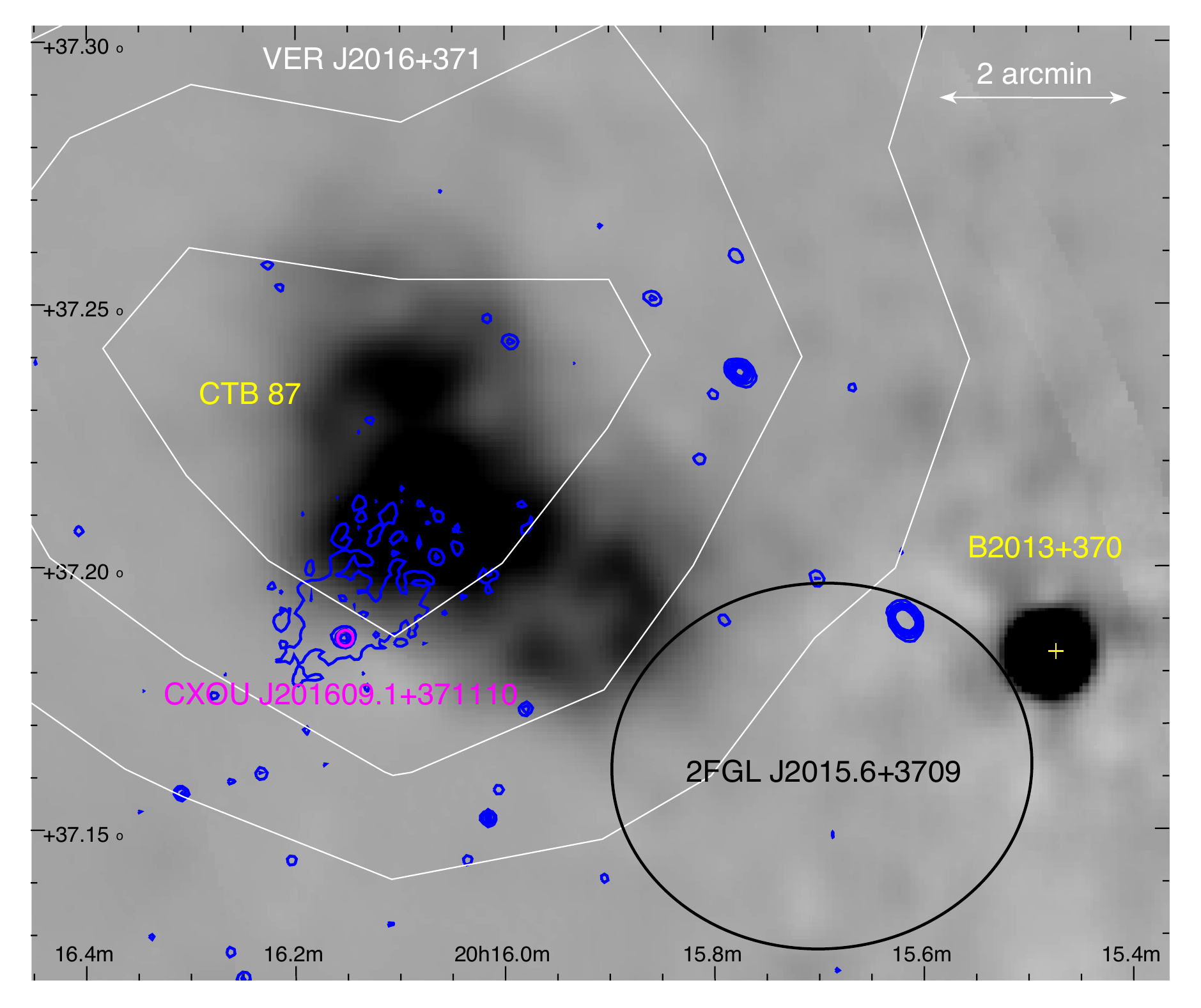}
\caption{ 610 MHz GMRT image of the CTB 87 region with X-ray and VHE gamma-ray contours in the equatorial coordinate system. The grey-scale image is smoothed with a 2D Gaussian ($\sigma=30\arcsec$). Blue contours indicate the X-ray morphology of the 0.3-7.5 keV Chandra image with the position of the putative pulsar indicated in magenta.  The VERITAS emission with integration radius of 0$^{\circ}$.089 is shown as overlaid white contours of significance 3, 4 and 5$\sigma$. The blazar B2013+370 and the {\bf 95\%} error ellipse of 2FGL J2015.6+3709, which are likely to be mutually associated, are also indicated. \label{CTB87a}}
\end{figure}

In Fig.~\ref{CTB87a} we present a false color image of the radio and X-ray emission in the region around VER J2016+371 obtained with the Giant Metrewave Radio Telescope ({\sl GMRT})~\citep{paredes09} at 610 MHz and {\sl Chandra} between 2 and 10 keV, respectively.~The VHE contours of VER J2016+371 are co-located with the bright and extended low-energy emission from the SNR CTB 87.~At radio wavelengths, the strong polarization, flat spectral index, center-filled morphology and lack of continuum shell have been used to classify CTB 87 as a PWN~\citep{weiler78,wallace97}.~The high angular resolution of the {\sl GMRT} image ($\sim30\arcsec$) shows a faint circular structure in the south-western portion of the nebula. Further studies at multiple wavelengths will be needed to determine if this structure is related to CTB 87 or perhaps a different source. The smoothed archival X-ray image reveals a centrally peaked morphology which is offset towards the south-east of the radio peak and has a slightly smaller extent than the radio emission.  The X-ray emission was recently studied in more detail by \citet{matheson13}. The superb angular resolution of {\sl Chandra} also allowed these authors to localize the pulsar candidate, CXOU J201609.2+371110, located within the compact PWN (to the south-east of the remnant center).  

HE gamma-ray emission is also detected in the vicinity of VER J2016+371 with the Large Area Telescope on board the {\sl Fermi} spacecraft~({\sl Fermi}-LAT)~\citep{abdo09b}. The 95\% error ellipse of the unidentified HE gamma-ray source 2FGL J2015.6+3709 does not exclude a common origin between the two sources. However, based on the variability index of the {\sl Fermi}-LAT source and its correlation with radio,~\citet{kara12} associate the HE gamma-ray emission with the nearby blazar B2013+370, with unknown redshift, rather than with the CTB 87. On the other hand, no VHE gamma-ray emission from this extragalactic object is seen in the current data. Its location lies $6.7\arcmin$ away from the centroid of VER J2016+371, this being much larger than the $\sim1.5\arcmin$ uncertainty of the VHE measurement.

\subsubsection*{A PWN scenario}

The morphology of the extended X-ray PWN~\citep{matheson13} suggests that it is affected by ram pressure due to the proper motion of the pulsar which must then be moving south-east. Since the radio emission is not coincident with the compact X-ray emission from the pulsar, the diffuse radio emission in the center of CTB 87 must be produced by an older generation of pulsar wind particles which have lost too much energy to emit X-rays via the synchrotron mechanism. A similar situation is seen for PSR B1706--44 in the SNR G343.1--2.3 (see Fig. 2 of~\cite{romani05}) which has been associated with the VHE source HESS J1708--443~\citep{hess2011}. We note that the morphology of the compact X-ray PWN in CTB 87 (again, similar to PSR B1706--44 in G343.1--2.3) implies that the pulsarÕs direction of motion is at a fairly large angle ($60^{\circ}-70^{\circ}$) with respect to the compact PWN symmetry axis (which is likely also the pulsar spin axis). 

If the pulsar was born at the apparent center of CTB 87 (i.e.\ $\sim2'$ away from the CXOU J201609.1+371110 position), the offset between the radio emission and the pulsar implies a transverse (projected onto the sky) velocity of $\simeq 70 d_{6.1} \tau_{50}^{-1} $ km s$^{-1}$ where $d_{6.1}=d/6.1~{\rm  kpc}$ is the distance of the SNR, derived from H I and CO observations~\citep{kothes03}, and $\tau_{50}=\tau/50~{\rm kyrs}$ an age scaled to an order of magnitude estimate based on ages of known ram pressure confined PWNe resolved in X-ray at the distances $>2$ kpc (see pulsars marked by asterisks  in Table 1 of \cite{kargaltsev13}.~Based on the fairly large X-ray luminosity, the pulsar could be younger than the reference age of 50 kyrs (see also \cite{matheson13}). In this case the pulsar velocity would become supersonic compared to the sound velocity of heated ejecta ($\sim 100$ km ${\rm s^{-1}}$), or highly supersonic if the pulsar has completely escaped the unseen SNR into the warm ISM, where the sound velocity is $\sim 10$ km ${\rm s^{-1}}$ (see e.g.,\cite{gaensler04}).

Alternatively, the offset between the radio emission and the X-ray nebula may largely be due to an asymmetric reverse shock that  pushed the relic radio emitting PWN away from the pulsar's current position. In both scenarios, the PWN needs to be old enough  for the reverse shock to have passed, making an age younger than 5-10 kyr unlikely.

In a commonly considered relic PWN scenario, where the X-rays are  attributed to synchrotron emission from pulsar wind and the VHE gamma rays are interpreted as the CMB photons up-scattered by pulsar wind electrons via the IC mechanism (see e.g., \cite{aharonian95}), the average magnetic field within the PWN can be estimated following arguments given in~\cite{aliu13} for a PWN in the CTA 1 SNR. These estimates lead to a somewhat high $B_{\rm PWN}\sim20-40~\mu$G, if diffusion is neglected for the pulsar wind particles, and to a more reasonable $B_{\rm PWN}\sim5~\mu$G, if diffusion is the dominant transport mechanism (see~\cite{aliu13} for details). The multiwavelength properties of VER J2016+371 are in line with those of other VHE PWNe~\citep{kargaltsev13};  therefore, VER J2016+371 is another example of a PWN which is seen in both X-rays and VHE gamma rays. \\
  


\begin{figure}[t]
\epsscale{1.3}
\plotone{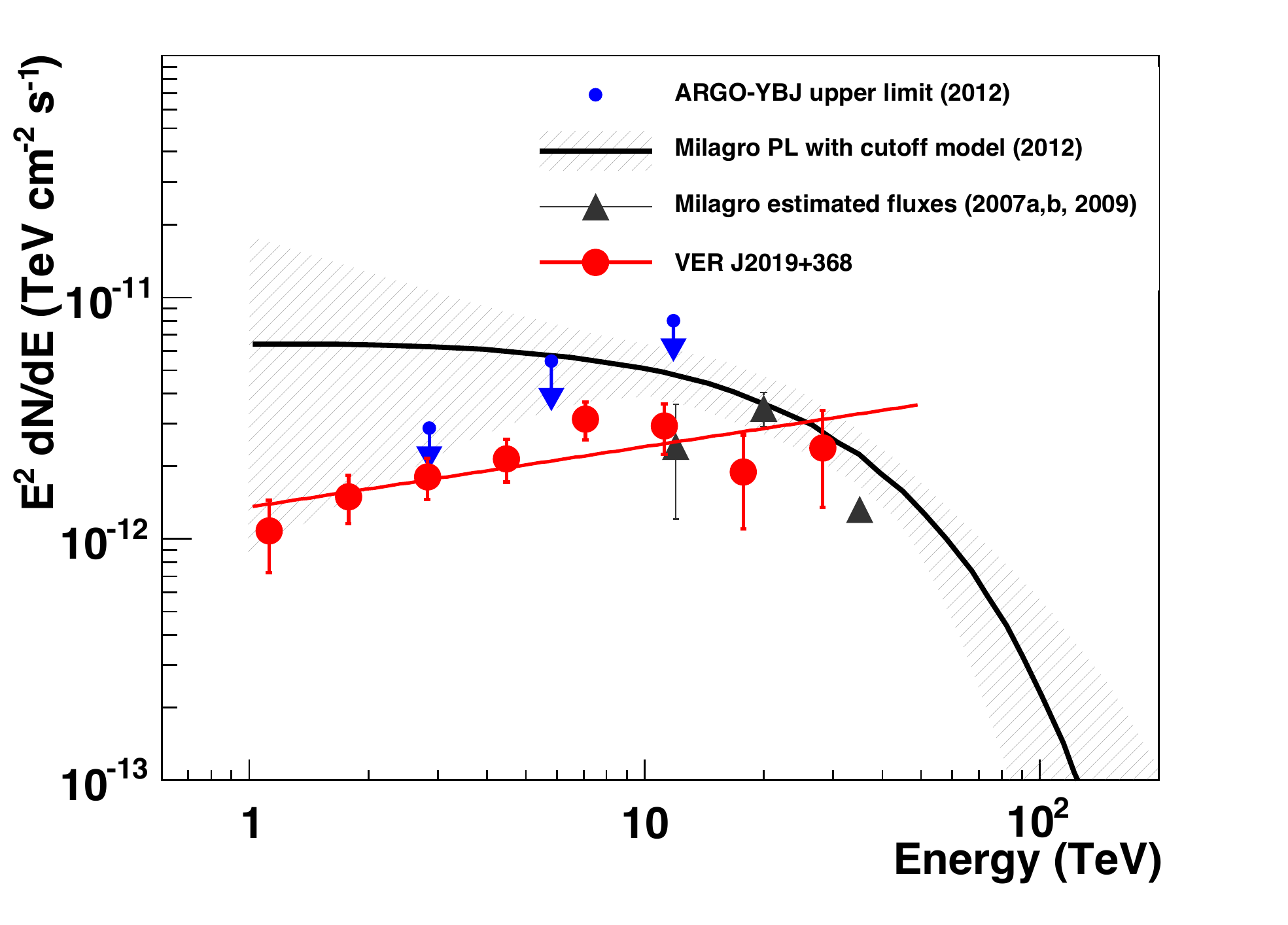}
\caption{Spectral energy distribution of MGRO J2019+37/VER J2019+368 as measured by different instruments. VERITAS measures a spectrum from 1 TeV to almost 30 TeV, shown in red, which is best fit with a power law with a hard spectral index. The Milagro flux points at 12, 20 and 35 TeV are shown in black~\citep{abdo07a, abdo07b, abdo09a} and also in black is their best fit, a power law with a cutoff~\citep{abdo12}. The shadowed area corresponds to the 1$\sigma$ band.~ARGO-YBJ 90\% confidence level upper limits for MGRO J2019+37 are shown with blue arrows~\citep{bartoli12}. \label{fig6}}
\end{figure}

\begin{figure*} [t!]
\epsscale{1.0}
\plotone{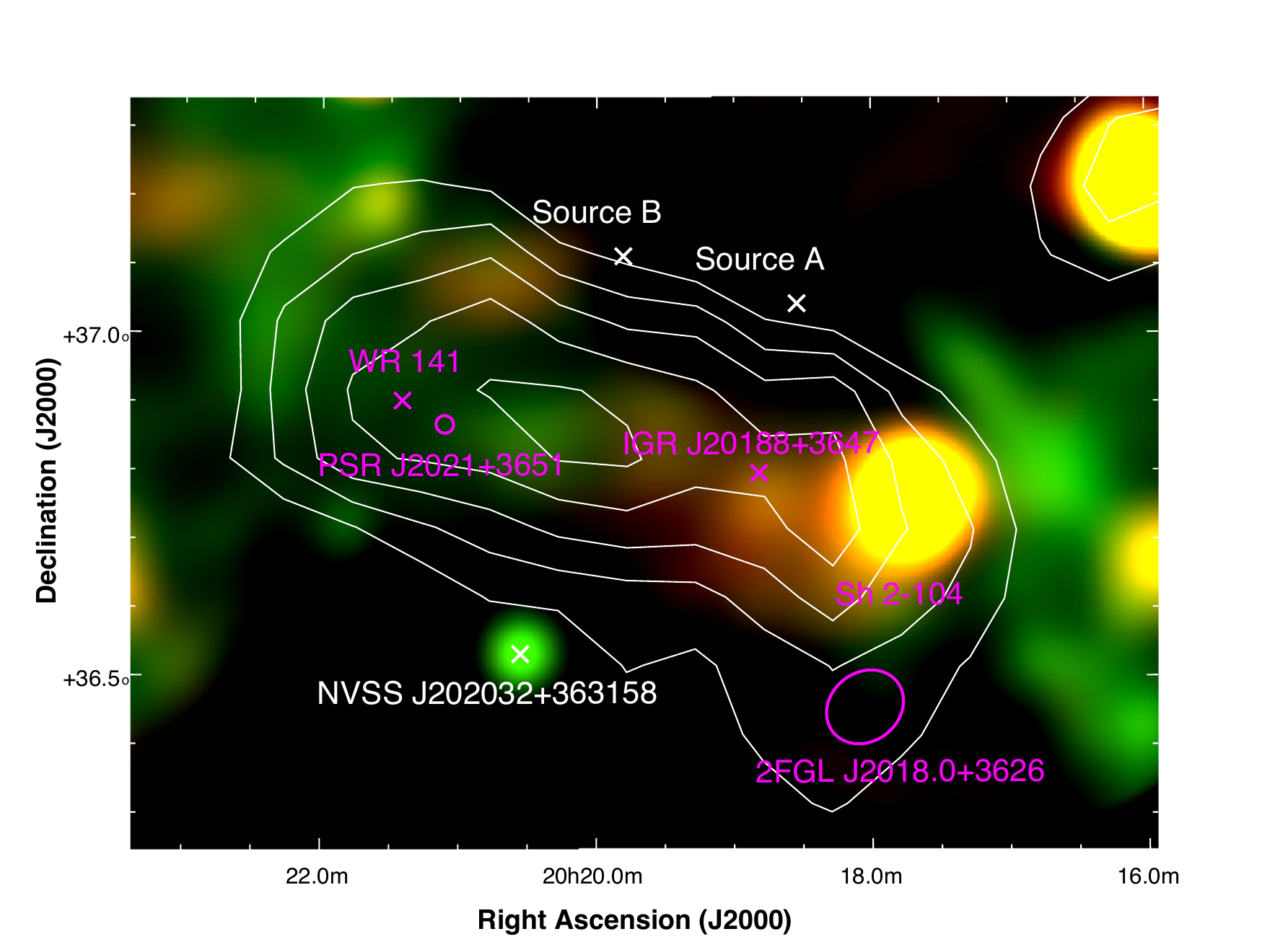}
\caption{ False color image of the radio diffuse emission of the VER J2019+368 region obtained with CGPS 408 MHz (green) and GB6 6cm (red). To produce this image we have taken the CGPS and GB6 radio images of the region and convolved them with the same beam size. The VERITAS significance contours at 3, 4, 5, 6 and 7$\sigma$ obtained with the large integration radius (0$^\circ$.23) are overlaid. Possible counterparts to MGRO J2019+37 within the literature are marked and labelled in magenta if they fall inside the 3$\sigma$ contours of VER J2019+368 and white otherwise. \label{fig7}}
\end{figure*}

\subsection{VER J2019+368, the main contributor of MGRO J2019+37}

VER J2019+368 is an extended source that is about four times brighter than VER J2016+371 at 1 TeV. The centroid of the VER J2019+368 emission is separated from VER J2016+371 by $\sim0^{\circ}.8$, and coincides well with the center region of MGRO J2019+37, as shown in Fig.~\ref{fig1}. The extension of VER J2019+368 is $\sim 0^\circ.35$ along the major axis, which is 50\% smaller than the extension of $0^{\circ}.7$ for MGRO J2019+37, as reported by \citet{abdo12}. Fig.~\ref{fig6} shows the spectral energy distribution (SED) of the MGRO J2019+37 region measured by different VHE gamma-ray instruments. ~\citet{abdo12} estimated the spectral index of MGRO J2019+37 to be 2.78 $\pm$ 0.1 for a PL hypothesis and $2.0_{-0.1}^{+0.5}$  for a PL with cut-off hypothesis. Their F-test favored the PL + cut-off model, for which the spectral index is in agreement with that measured by VERITAS for VER J2019+368, 1.75 $\pm$ 0.38. The spectrum of VER J2019+368 can be explained by a PL model up to $\sim$ 30 TeV. Even though the flux from CTB 87 is not included for VERITAS measurements (unlike Milagro's measurements including the emissions from a larger region including CTB 87), the spectra are consistent. As seen in Fig.~\ref{fig6}, the flux of VER J2019+368 is consistent with fluxes estimated by Milagro at 12, 20 and 35 TeV and is also in agreement with the upper limit estimated by ARGO-YBJ. Based on the consistent flux levels and coincident location of the centroids between the two measurements, we expect VER J2019+368 to be the main contributor to the VHE emission from MGRO J2019+37.  

Several surveys in radio, infrared, and X-ray wavelengths have studied the inner and brighter region of MGRO J2019+37 in an attempt to identify potential counterparts to the VHE emission. Since VER J2019+368 likely shares the counterparts of the inner MGRO J2019+37 and provides better localization, we are now able to reevaluate the possible sources and mechanisms generating the VHE emission. The potential counterparts suggested from the surveys can be found in Fig~\ref{fig7}, marked over the radio continuum images of the region. These radio images are from the Dominion Radio Astrophysical Observatory (DRAO) within the Canadian Galactic Plane Survey (CGPS) project at 408~MHz~\citep{taylor03} and the Green Bank Telescope at 4.85 GHz (GB6). The VHE emission appears to follow a ridge of diffuse emission starting at the bright bubble H{\sc ii} region Sh 2-104 at the west end and roughly ending near the energetic gamma-ray pulsar PSR J2021+3651.

\subsubsection{WR 141}
 This Wolf-Rayet - O star colliding wind binary system is in the Cygnus arm at a distance of $\sim 1.3$~kpc \citep{v01}. The estimated terminal wind velocity and mass loss rates from the system suggest it is dumping energy into its surroundings at a rate of $\sim 2\times 10^{37} {\rm erg}\,{\rm s^{-1}}$. While high-energy gamma-ray emission is theoretically expected to arise from such a system, ~\cite{reimer06} predicts a suppression of the highest energetic photons in the VHE band in the Klein-Nishina regime. In addition, there is nothing particularly remarkable about the X-ray, radio and optical studies of WR 141 compared to similar systems that are not associated with bright VHE emission \citep{z12,mpa+09,mme98}, which makes this object a less likely counterpart for producing the observed VHE gamma rays.\\
 
\subsubsection{PSR J2021+3651 and its PWN}

Located at $\sim 20'$ eastward from the centroid of VER J2019+368, the young and energetic pulsar PSR J2021+3651 has a high spin-down luminosity ($\dot{E}=3.4\times10^{36}$ erg/s;~\citet{roberts02}). X-ray observations with {\sl Chandra} revealed a compact X-ray nebula with a clearly resolved torus and jet morphology surrounded by extended, diffuse emission~\citep{hessels04,vanetten08}. According to the X-ray and VHE PWN population study by \citet{kargaltsev10}, typically only pulsars with spin-down powers higher than $10^{35}$ erg/s are associated with prominent VHE PWNe. The high spin-down luminosity of PSR J2021+3651 makes plausible a VHE PWN scenario for VER J2019+368.

\begin{figure} [t]
\epsscale{1.1}
\plotone{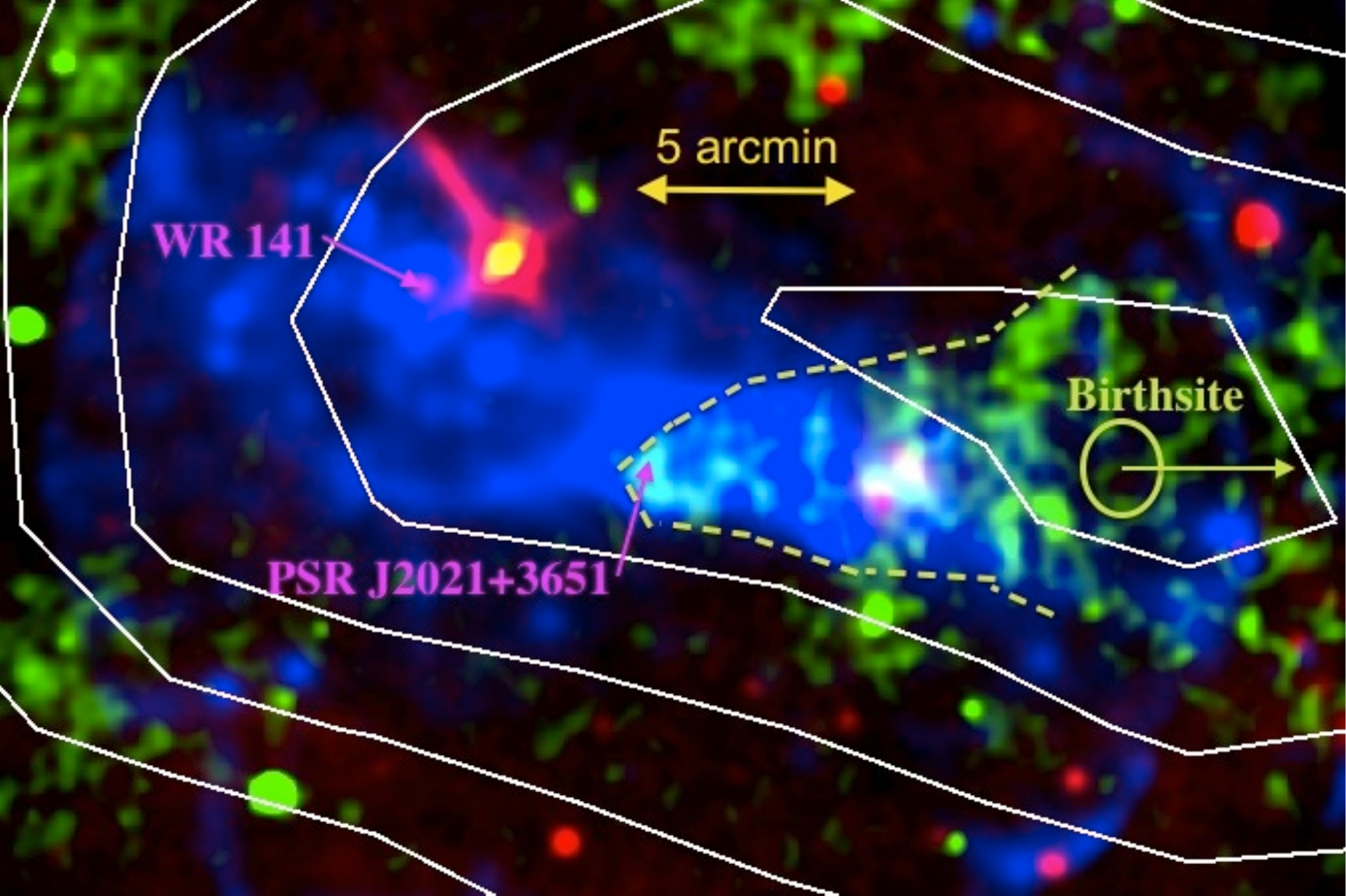} 
\caption{False color image of the infrared, radio and X-ray emission in the vicinity of PSR J2021+3651 obtained with MSX 8.3$\mu m$ (red), VLA 20cm (green) and XMM 1-8 keV (blue), respectively. The VERITAS significance contours at 3, 4, 5, 6 and 7$\sigma$ obtained with the large integration radius (0$^\circ$.23) are also shown. Overlaid dashed line highligths the PWN morphology in radio, and an ellipse marks the closest birthsite suggested by this morphology. An arrow indicates the direction of the birthsite if happened further away.\label{fig8}}
\end{figure}

To understand the overall picture, we performed a deep observation of the PSR J2021+3651 region with $XMM-Newton$ (PI Roberts) which was combined with archival observations for a total effective exposure (after removing times of high background) for each MOS detector of $\sim 105$~ks. The complexity of the extended X-ray emission overlaid with infrared and radio emission is shown in Fig.~\ref{fig8}. A detailed analysis of these data will be presented elsewhere. Here we note that multi-color imaging (infrared, radio and X-rays) shows that the extended emission centered on the pulsar is almost all non-thermal, and there is little evidence for strong spectral variation. The morphology of the X-ray emission to the west of the pulsar is that of a brighter tail of emission within a roughly bow shaped nebula that conforms to the shape of the radio nebula. With the additional evidence of the same approximate shape from the outer X-ray nebula to the west of the pulsar, we can plausibly assume that the cone-shaped region seen in the VLA image in Fig.~\ref{fig8} starting at the pulsar and extending ~10$'$ west
belongs to the PWN of PSR J2021+3651.

If the shape is due to the pulsar's motion, then it is likely the pulsar was born at least as far west as the apparent end of the radio tail. Given the pulsar's characteristic age of $\tau_c \sim 17000$~yr, this implies a transverse motion of $\sim  0.035^{{\prime \prime}} \, {\rm yr}^{-1}$. 
If we knew the distance to the pulsar, then we could estimate its transverse velocity. Unfortunately, the distance to PSR J2021+3651 is controversial. The dispersion measure is quite large, which yields an estimated distance of $\sim 12$~kpc from the NE 2001 free electron model \citep{cordes02}.~\citet{roberts02} pointed out that such a large distance would imply a very high $\gamma$-ray efficiency compared with other known pulsars, and a PWN X-ray efficiency at the high end of the distribution for young pulsars. Considering the X-ray emission and absorption, as well as excess scattering of the radio pulse indicating line of sight material not accounted for in the NE2001 model, \citet{hessels04} suggested a distance of  $\sim 8$~kpc might be more reasonable. Other authors argued that, despite the large radio dispersion measure, the nebular size and the thermal X-ray emission from the pulsar surface indicated a distance as low as $4$~kpc~\citep{vanetten08}. However, the sense and magnitude of the polarization rotation measure seem to imply a minimum distance of $\sim 5$ kpc \citep{abdo09b}.
Assuming a distance of 5 kpc, the implied transverse velocity of the pulsar is $v=840d_5{\tau}_{17}^{-1} {\rm km}\,{\rm s}^{-1}$, where $\tau_{17} = \tau/17000{\:\rm yr}$ and $d_5 = d/5{\:\rm kpc}$. This velocity is 3-4 times higher than average for pulsar proper velocities, but would not be record breaking \citep{chatt05}.
Since the true age is very unlikely to be greater than twice the characteristic age~\citep{kaspi01}, the pulsar most likely could not have been born much farther west than the extent of the radio tail (Fig~\ref{fig8}) unless its actual distance from Earth is much smaller than $\sim5$ kpc.
In particular, it is unlikely that the pulsar was born near the west edge of the VHE emission, which is an additional $\sim 30^{\prime}$ further west and makes it less likely to be responsible for the entire VHE emission.  

X-ray measurements of PSR J2021+3651 reveal that its properties are similar to those of the nearby young pulsar Vela-X. Both pulsars have high spin-down luminosities (the spin-down luminosity of Vela-X is twice that of PSR J2021+3651) and possess compact inner toruses~\citep{roberts02,vanetten08}. Also, the remarkably hard spectrum of VER J2019+368 is similar to the hard TeV emission that was observed in Vela-X \citep{abramowski12}. Although a single VHE PWN scenario may not be plausible to explain the entire VHE emission, based on the distance and velocity estimation, emission from the PWN is likely responsible for a significant portion of the emission.

\subsubsection{IGR J20188+3647}
The location of the hard X-ray transient IGR J20188+3647 lies near the maximum of the VHE emission. This transient was reported on July 2004, and its flaring behavior consisted of a fast rise ($\sim$10 minutes) followed by a slower decay ($\sim$50 minutes)~\citep{sguera08}. Following a flare detected by AGILE in the region, a $\sim$21 ks {\sl XMM-Newton} observation took place on November 2007 around IGR J20188+3647~\citep{zabalza10}. In that observation, five X-ray candidate sources appear inside the error box of the INTEGRAL source of $3.4'$. Four of them are associated with late stellar types, while the fifth one, which appears to be a highly absorbed hard X-ray source, is considered to be either an AGN or a pulsar. We carried out observations with the Swift X-ray telescope ({\sl Swift}-XRT) spread over three months (16 th June, July and August 2011, PI Aliu), aiming at finding some brightening of these X-ray sources to identify a potential counterpart for this transient, but no likely counterparts were found. 

\subsubsection{Sh 2-104}
The western edge of VER J2019+368 overlaps the H{\sc ii} region Sh 2-104. At a distance of $\sim 4$~kpc, Sh 2-104 is beyond the Cygnus-X region. Sh 2-104 has a shell morphology in both optical and radio wavelengths with an O6V star in the center, which is ionizing the region. ~At the eastern edge of the shell, there is an ultracompact H{\sc ii} region, coincident with stellar clusters \citep{deharveng03}. 
With massive CO clouds around the star cluster, \cite{deharveng03} suggested that the Sh 2-104 region is a prototype of massive-star formation triggered by the expansion of an H{\sc ii} region. Star forming regions and stellar clusters are considered to be possible gamma-ray sources, with the gamma rays arising from shocks created by the wind of single or multiple massive stars colliding with the surrounding material \citep{torres04}. We cannot rule out the contribution from Sh 2-104 because of its nearby location to the VHE emission, although $\sim 6000 M_{\odot}$ of swept up mass of Sh 2-104 seems to be low compared to other star forming regions that are associated with VHE gamma-ray sources, such as W49A\citep{brun10} or Westerlund1\citep{luna10}. 


\subsubsection{2FGL J2018.0+3626}
An unassociated {\sl Fermi}-LAT source, 2FGL J2018.0+3626, appears towards the south-west of VER J2019+368. This source is non-variable and shows a steep spectrum with a cutoff around 4 GeV, thus exhibiting properties similar to those of gamma-ray pulsars. Because the distance between 2FGL J2018.0+3626 and the centroid of VER J2019+368 is relatively large, it is unlikely that this source is directly responsible for the major part of the VHE emission. However, the softening at the west side of VER J2019+368 shown in Fig.~\ref{fig4} could be associated to this secondary contribution.  If the 2FGL source turns out to be an energetic pulsar, then it is possible for a PWN to contribute part of the west side of the VHE emission. To check this scenario, identifying the nature of 2FGL J2018.0+3626 is necessary. Unfortunately, the region is not well observed in X-rays, making a search for a possible pulsar or a PWN difficult. The position of  2FGL J2018.0+3626 is outside the FOV of the {\sl XMM-Newton} search reported by \citet{zabalza10}. The low resolution radio surveys suggest some possible faint emission, but it is hard to quantify because of nearby bright emissions from Sh 2-104.


\subsubsection{Other suggested counterparts}
 \citet{paredes09} searched the region for possible counterparts to the Milagro emission by using {\sl GMRT} 610~MHz data. Together with  PSR J2021+3651 and the H{\sc ii} region Sh 2-104 that we mentioned above,~\citet{paredes09} suggested the possibility of contributions from three non-thermal sources with jet-like structures. The improved VHE image of the region obtained with VERITAS excludes these three, labeled as sources A, B and NVSS J202032+363158, which are marked in white in Fig.~\ref{fig7}. These sources fall outside the 3$\sigma$ emission region from VERITAS.

\section{Summary and Conclusion}
We have carried out a deep VHE observation in the region of MGRO J2019+37 with VERITAS, confirming this to be a very gamma-ray rich area of the Galactic Plane with emission detected also with {\sl Fermi}-LAT and Milagro, and, thus,  covering the full range of the high energy spectrum. 

The new VHE image spatially resolves, for the first time, one Milagro extended source into at least two clearly separate sources. The angular resolution is good enough to exclude some scenarios that were proposed for MGRO J2019+37.  The most likely counterpart of the new source VER J2016+371 is the PWN in the SNR CTB 87. The co-location, the VHE extent and the X-ray/VHE luminosity ratio argue in favor of this. More complicated is the multi-wavelength emission from VER J2019+368, as there are a few possible explanations for its physical origin.  The young and energetic pulsar PSR J2021+3651 and its PWN, proposed by many before,  seems a likely contributor to VER J2019+368, and therefore MGRO J2019+37. The extended VHE morphology in the direction of the X-ray and radio nebula favors this possibility. The very hard spectrum of the source, $\Gamma=1.75\pm0.08_{stat}\pm0.3_{sys}$, which resembles that of Vela X~\citep{aharonian06}, another PWN system, also favors this scenario. However, the picture might be more complex as non-thermal X-ray emission coincident with Sh 2-104  and also an unassociated {\sl Fermi}-LAT source are also in physical association with this large VHE emission. Follow-up multi-wavelength observations of the region, including VHE data,  will certainly help clarify which scenario or scenarios are in fact originating VER J2019+368. 


\acknowledgements
The authors would like to thank Jules Halpern for many useful discussions and help in obtaining and interpreting the Swift data.This research is supported by grants from the U.S. Department of Energy Office of Science, the U.S. National Science Foundation and the Smithsonian Institution, by NSERC in Canada, by Science Foundation Ireland (SFI 10/RFP/AST2748) and by STFC in the U.K. We acknowledge the excellent work of the technical support staff at the Fred Lawrence Whipple Observatory and at the collaborating institutions in the construction and operation of the instrument. Based on observations obtained with XMM-Newton, an ESA science mission with instruments and contributions directly funded by ESA Member States and NASA. This research has made use of data and/or software provided by the High Energy Astrophysics Science Archive Research Center (HEASARC), which is a service of the Astrophysics Science Division at NASA/GSFC and the High Energy Astrophysics Division of the Smithsonian Astrophysical Observatory. The research presented in this paper has used data from the Canadian Galactic Plane Survey, a Canadian project with international partners, supported by the Natural Sciences and Engineering Research Council.

{\it Facility:} \facility{VERITAS}.

\end{document}